\documentstyle[12pt,epsfig]{article}

\newcommand{\be}{\begin{equation}}
\newcommand{\ee}{\end{equation}}
\newcommand{\bea}{\begin{eqnarray}}
\newcommand{\eea}{\end{eqnarray}}
\begin{document}

\begin{center}
{\Large \bf Beyond the soft photon approximation in radiative production
and decay of charged vector mesons.}\\

\vspace{1.3cm}

{\large G. L\'opez Castro}\\

{\it Departamento de F\'{\i}sica CINVESTAV-IPN
A.~P. 14-740 M\'exico 07000 D.F. M\'exico}\\

{\large  G. Toledo S\'anchez}.\\

{\it Department of Physics, Florida State University,
Tallahassee, FL 32306. USA}
\end{center}
\vspace{1.3cm}

\begin{abstract}

We study the effects of model-dependent contributions and the electric
quadrupole moment of vector mesons in the decays $V^- \rightarrow
P^-P^0\gamma$ and $\tau^- \rightarrow \nu V^-\gamma$.
Their interference with the amplitude originating from the
radiation due to electric charges vanishes for photons emitted collinearly
to the charged particle in the final state.
This brings further support to our claim in previous works, that
measurements of the photon energy spectrum for nearly collinear photons in
those decays are suitable for a first measurement of the magnetic dipole
moment of charged vector mesons.

\end{abstract}
 
  In some recent papers \cite{us97,us99,us00}, we have studied the
possibility to obtain information about the experimental value of the
magnetic dipole moment (MDM) of light charged vector mesons.
We have focused our interest on the angular and
energy distribution of photons emitted during the radiative decay
\cite{us97} or production \cite{us99} processes of vector mesons. In Ref.
\cite{us00} we have extended our analysis to include the finite width
effects of these unstable particles. Just  to illustrate the interest of
the problem, let us emphasize that none of the magnetic dipole moments of
resonances, neither elementary (as the W gauge boson) nor hadronic
particles, have been measured yet. 
The only exception is the $\Delta^{++}$ resonance, where the  
determinations of the MDM spread over a wide range \cite{pdg2000}: $3.5
<\mu_{\Delta^{++}} < 7.5$ in units of nuclear magnetons. This
particular case makes inconclusive any comparison with theoretical
predictions based on quark models \cite{pred}. Therefore, observables
that could provide information about the MDM of unstable particles, open
up the possibility to test additional features of the dynamics of bound
states in strong interactions.

  The  energy spectrum of  photons emitted at small
angles with respect to final state charged particles in the decays $V^-
\rightarrow P^-P^0\gamma$ \cite{us97} and $\tau^- \rightarrow \nu
V^-\gamma$ \cite{us99} ($V(P)$ denotes a vector (pseudoscalar) meson),
was found to be particularly sensitive to the effects of the MDM of the
vector meson. We should recognize, however, the challenges that
reconstructing these channels from multi-photon final states may pose to
experimentalists. 

In our analysis of Refs. \cite{us97,us99,us00} we have neglected
model-dependent  
contributions and the effects coming from the electric quadrupole
moment of the vector meson.  We have argued, based on the fact that their
lowest order contributions in the photon energy vanish for collinear
photons, that these effects are expected
to be small for the particular kinematical region studied in those papers
\cite{us97,us99}.
In order to have a quantitative estimate of these effects,
in the present paper we focus  on the details of our proof that they are
indeed small compared to the dominant effects coming from the MDM of the
vector meson.\\

Let us first discuss the general structure of the decay amplitude for
the emission of a low-energy photon. The decay amplitude for the
process $i \rightarrow f+\gamma$ (either, $i$ and/or $f$ states containing
charged particles) can be expressed as a power expansion in the photon
energy $\omega$ as follows \cite{low}:

\be
{\cal M} =\frac{A}{\omega} +B\omega^0 + C\omega +\cdots 
\ee

Low's theorem \cite{low} states that the first two terms of this
amplitude are model-independent. This means that the coefficients $A$ and
$B$ can depend only on the parameters that describe the corresponding
non-radiative process $i \rightarrow f $ and on the static
properties (electric charges and
magnetic dipole moments) of the particles involved in $i$ and $f$.
Furthermore, the piece of the amplitude involving terms up to order
$\omega^0$ (also called Low's amplitude) must be gauge-invariant. The
electric charges contribute to the amplitude at order $\omega^{-1}$ while
the effects of the MDM enter at order $\omega^0$. In the rest frame
of decaying particles, Lows's amplitude is expected to give a very good
approximation to the radiative process for photon energies below a typical
 energy scale of the final particles (let's say, the pion mass for the
$\rho^-$ meson decay considered below).

   In equation (1), the terms of order one and higher in $\omega$
involve, in general, model-dependent contributions and the effects coming
from the electric
quadrupole moment of vector mesons\footnote{In this
paper we will consider decay process involving charged spin 1 particles. 
We will assume exact CP symmetry, which implies the vanishing of the 
electric dipole  and magnetic quadrupole moments of the vector
particles.}. Since Low's amplitude is gauge-invariant, the terms that
enter the amplitude (1) at  ${\cal O}(\omega)$  must be  gauge-invariant
by themselves. On another hand, by including {\it all} the model-dependent
contributions in Eq. (1) one would expect to obtain an amplitude valid for
the whole range of photon energies. In practice (as is the case of the
present work), one is limited to include only a few sources of
dominant model-dependent contributions. The final judgment on the
validity of these approximations corresponds to the experiment.

 The squared amplitude obtained after summation over polarization of
particles in $i$ and $f$, exhibits other
interesting properties. This unpolarized probability has the following
structure in terms of the expansion in  $\omega$:

\be
\sum_{pols \ i, f} |{\cal M}|^2 =  \frac{\alpha_2}{\omega^2}+ \alpha_0
\omega^0 + O(\omega)\ ,
\ee
{\it i. e.},  the interference terms of order $\omega^{-1}$ vanish. This
is the well known Burnett and Kroll's theorem \cite{bk} (see also
\cite{zkp}). 

   As it will be shown below, $\alpha_2$ is proportional to $\sin^2 \theta$
($\theta$ being the angle of the photon emission with respect to a given
charged particle of the process) and it comes from the radiation off
external charged particles. On another hand, $\alpha_0$ receives two
contributions: the term proportional to $|B|^2$ and the interference
term Re $\{ AC^*\}$. For the cases of our interest here, the
angular dependence of the $|B|^2$ term is very mild as it can be
appreciated from the plots in our previous papers \cite{us97, us99}.
Therefore, the decay probability (2) for $i \rightarrow
f+\gamma$ at order $\omega^0$ is in general model-dependent. A
reliable information about the magnetic dipole moment can be obtained
provided the contributions coming from model-dependent and electric
quadrupole moment effects are small. This can be explicitly proven for radiative decay and production of
charged vector mesons. In the following we will show that the interference
term Re$\{AC^*\}$ is also suppressed for the kinematical configurations
chosen
in  Refs. {\cite{us97,us99}. We divide  our
analysis into two parts: first we consider the decay  $V^- \rightarrow
P^- P^0 \gamma$ and then we focus on the production $\tau^- \rightarrow
V^-
\nu \gamma$ processes of charged vector mesons $V^-$ ($P$ denote a
pseudoscalar meson).\\

Let us first introduce some useful notation.  Let  $q$ and
$p$ be the four-momenta of the decaying and the charged particle in
the final state, respectively, and $k\ (\epsilon)$ the four-momentum
(polarization) of the photon. We define the four-vector:
\be
L^\mu \equiv \left( \frac{p^\mu}{p\cdot k} - \frac{q^\mu}{q\cdot k}\right)
\ee
which satisfies the  property $L\cdot k=0$. In the rest
frame of the decaying particle, we have 
\be 
L^2=L \cdot L =-\frac{|\vec{p}|^2}{(p\cdot
k)^2} \sin^2\theta
\ee
where $\theta$ is the angle between the photon three-momentum and 
$\vec{p}$.\\

\begin{center}
{\bf Vector meson decays: $V^- \rightarrow P^-P^0 \gamma$ \\}
\end{center}

We first consider the decay process of vector mesons and, for
definiteness, we choose the radiative decay $\rho^- (q,\eta) \rightarrow
\pi^- (p) \pi^0 (p') \gamma(k,\epsilon)$ (Latin and Greek letters within
parenthesis denote four-momenta and  polarization four-vectors,
respectively). The
Low's amplitude can be written as \cite{us97}:

\begin{eqnarray}
{\cal M}_L&=&ieg  (p-p')\cdot\eta L\cdot \epsilon^* \\
&&+ieg(k\cdot\eta\frac{q\cdot\epsilon^*}{q\cdot
k}-\epsilon^*\cdot\eta)F_1
+ieg(k\cdot\eta\frac{p\cdot\epsilon^*}{p\cdot k}-\epsilon^*\cdot\eta)F_2\
, \nonumber
\end{eqnarray}
where
$F_1 \equiv
\frac{m_\pi^2-m_\pi^{'2}}{m_\rho^2}(1-\frac{\beta(0)}{2})-\frac{\beta(0)}{2}$, 
$F_2 \equiv -1+\frac{\beta(0) p\cdot k}{q\cdot k}$, and $g \simeq 6.01$
denotes the
$\rho \pi\pi$ coupling, $e$ is the electric charge of the positron,
$m_\rho$ is the $\rho$ meson mass and $m_\pi$ ($m'_\pi$) is the mass of
the charged (neutral) pion.  The first term of this amplitude,  hereafter
called ${\cal M}_e$, is proportional  to $L$ and can be identified with
the term $A/\omega$ in Eq. (1). This term is gauge invariant by itself and
$\sum_{pols}|{\cal M}_e|^2$, is proportional to $L^2$. The remaining terms
of order $\omega^0$ in Eq. (5) are also gauge invariant, contain the
effects of the magnetic dipole moment $\beta$(0) (in units of $e/2m_{\rho}$),
and can be identified with the term $B$ in Eq. (1). $\beta(0)=2$,
corresponds to the {\it canonical} value for point particles.

Now let us consider the model-dependent contributions to the term $C$ in
Eq. (1). They are dominated by those terms coming from intermediate  
mesons through the sequence 
$\rho^- \rightarrow \pi^- X^0 \rightarrow \pi^-\pi^0\gamma$ where
$X^0$ is an isoscalar vector meson ($\omega,\phi$) or a neutral  axial
meson ($a_1$). 
The contribution from the $a_1^0$ meson is not possible because C parity,
while that
from the $\phi$ meson is suppressed by the
Okubo-Zweig-Iizuka rule. Thus, we
are left with the $\omega$ vector meson as the dominant identified source
of model-dependent contributions \footnote{The contribution of the
intermediate $\omega$ meson to the
branching ratio of $\rho^- \rightarrow \pi^-\pi^0\gamma$ lies two orders
or magnitude below the model-independent contribution given  by Low's
amplitude (see Bramon et al in Ref. \cite{us97}.)}. 

The contribution to the amplitude coming from the $\omega$ meson  
intermediate state is given by:

\be
{\cal M}_d=-\frac{ g_{\omega\pi^0 \gamma} g_{\rho \omega\pi}}
{m^2_\rho + m^2_\pi - m^2_\omega-2q\cdot p}
\epsilon^{*\nu} \eta^\theta  \epsilon_{\mu\nu\alpha\delta}
\epsilon_{\lambda
\theta\tau\delta} k^\mu (q-p)^\alpha q^\lambda (q-p)^\tau
\ee
where  $m_\omega$ is the mass of the $\omega$ meson and $g_{\omega\pi^0
\gamma}\simeq  0.811\ GeV^{-1}$, $g_{\rho \omega\pi}\simeq 13.5 \ 
GeV^{-1} $ are 
their corresponding coupling constants to final and initial states,
respectively. Note
that this amplitude is  gauge-invariant  and of ${\cal O}(\omega)$ in the
photon energy.

 The interference of this amplitude with the term of ${\cal
O}(\omega^{-1})$ in the Low's
amplitude, ${\cal M}_e$, summed over polarizations is 

\be
\sum_{pol}{\cal M}_d{\cal M}^*_e =-\frac{2eg g_{\omega\pi^0 \gamma}
g_{\rho\omega\pi} k\cdot p (k\cdot q)^2}
 {m^2_\rho + m^2_\pi - m^2_\omega-2q\cdot p}L^2.
\ee
Thus, the contribution of model-dependent terms, at their leading order
in $\omega$, will be suppressed for small angles of photon emission since
they are proportional to $L^2$.

We focus now on the contribution of the electric quadrupole moment of the
$\rho$ meson to the
term $C$ in Eq. (1). This multipole enters the electromagnetic vertex
$\rho^-(q) \rightarrow \rho^-(q')\gamma(k)$ as follows \cite{sakita}
(Lorentz indices $\nu\ (\lambda)$ refers to incoming (outgoing)) $\rho^-$
meson and $\mu$ to the photon external lines):
 \be
ie\Gamma_Q^{\mu\nu\lambda}= ie\gamma(0)\{ (2q-k)^\mu k^\nu k^\lambda
-q.k (k^{\lambda}g^{\mu\nu}+k^{\nu}g^{\mu\lambda}) \}\ ,
\ee
where $\gamma(0)$ denotes the electric quadrupole moment
\footnote{Quark model predictions \cite{pred} for the  quadrupole
moment cluster around 0.055 $fm^2$.  We will use a value $\gamma(0)=2$ in
our numerical estimations which corresponds to 0.065 $fm^2$, {\it i.e.}
not very far from theoretical expectations.} in
units of $e/2m_{\rho}^2$. Note that  $k_{\mu}\Gamma_Q^{\mu\nu\lambda}=0$,
namely the vertex is gauge-invariant on its own.

Its contribution to the decay amplitude is given by the gauge-invariant
term:
\be
{\cal M}_{Q}=-\frac{ieg \gamma(0)}{2} \left \{ (2p -q)\cdot k \left (
\frac{q\cdot \epsilon}{q\cdot k}k\cdot \eta-\eta \cdot \epsilon \right) -
2k\cdot
\eta p\cdot k L\cdot \epsilon \right\}\ .
\ee
 The interference term coming from ${\cal M}_{Q}$ and ${\cal M}_{e}$ is:
\be
\sum_{pol}{\cal M}_{Q}{\cal M}^*_e=-(eg)^2\gamma(0)\left\{ 
2 - \frac{q\cdot k}{m_\rho^2}\right\} p\cdot k(2p -q)\cdot k \L^2\ .
\ee

 In summary, since interference terms of order $\omega^0$ coming
from model-dependent contributions and from the electric quadrupole 
moment behave as
 \be
2{\rm Re}(AC^*) ={\rm Re} \{{\cal M}_e \cdot ({\cal M}_d^*+{\cal M}_Q^*)\}
\propto L^2\ ,
\ee
and using its proportionality to $sin^2(\theta)$ ($\theta$ the angle
between the corresponding photon and  charged pion 3-momentum), we
conclude that model-dependent and quadrupole moment contributions to
$\rho^- \rightarrow \pi^-\pi^0\gamma$, at leading order in $\omega$ are
suppressed when $\pi^-$ and $\gamma$ are emitted nearly to the collinear
configuration.\\

\begin{center}
{\bf Production of vector mesons: $\tau^- \rightarrow V^- \nu \gamma$ \\}    
\end{center}

Let us consider the decay $\tau^-(q) \rightarrow \rho^-(p,\eta)
\nu(p')\gamma(k,\epsilon)$ as an example of radiative production of
charged vector mesons \cite{us99}. The corresponding 
Low's amplitude is given by \cite{us99}

\begin{eqnarray}
{\cal M}_L &=& \frac{eg_{\rho}G_F V_{ud}}{\sqrt{2}}
\left\{
 \overline{u}(p') O^\alpha u(q)
\eta^*_\alpha L\cdot \epsilon^* 
\right. \label{tau} \\
&& + \left. 
 \overline{u}(p') O^\alpha \frac{\not{k}
\not{\epsilon}^* }{2q \cdot k}u(q) \eta_\alpha +
 \overline{u}(p')
\frac{(1+\gamma_5)}{p \cdot k} 
G^\alpha u(q)
(k_\alpha \epsilon^*_\lambda- \epsilon^*_\alpha k_\lambda) \eta^\lambda
\right \}. \nonumber
\end{eqnarray}
where $G^\alpha \equiv \frac{\beta(0)}{2} \gamma^\alpha
+(1-\frac{\beta(0)}{2})\frac{m_\tau}{m^2_\rho} p^\alpha$,
$G_F$ is the Fermi constant, $V_{ud}$ the Cabibbo-Kobayashi-Maskawa
$ud$ mixing, $g_{\rho} \simeq 0.166$ GeV$^2$ the $\rho-W$ coupling, 
and $O^\alpha \equiv \gamma^\alpha(1-\gamma_5)$.
As in the case of vector meson decays, the first term of this amplitude
(${\cal M}_e$), of order
$\omega^{-1}$, is proportional to the four-vector $L$ and identified with
the electric
charge radiation amplitude. Thus, its associated probability 
will be suppressed for small values of $\sin \theta$ ($\theta$ here is
the angle between the three-momenta of the $\rho^-$ and the photon in the 
rest frame of the $\tau$ lepton). The second and third terms
 are of $O(\omega^0)$ and correspond to the term $B$ in Eq. (1) .\\

  In the following we consider the contributions of order $\omega$ to the
decay amplitude of $\tau^- \rightarrow \rho^- \nu \gamma$. First, we can
identify the model-dependent contribution given by the $\pi^-$
intermediate state:  $\tau^- \rightarrow \pi^- \nu \rightarrow \rho^- \nu 
\gamma$. This contribution would be expected to provide the dominant
model-dependent channel,  given the fact that couplings of other hadrons
to the weak charged current are strongly suppressed.

   The corresponding gauge-invariant decay amplitude is given by:
\be
{\cal M}_\pi= \frac{iG_F f_\pi g_{\rho\pi\gamma}
}{\sqrt{2}((p+k)^2-m_\pi^2)}
 \epsilon_{\nu\mu\alpha\lambda} k^\nu \epsilon^\mu p^\alpha
\eta^\lambda
\bar{u}(p') O^\beta u(q) (p+k)_\beta \ ,
\ee
where $g_{\rho\pi\gamma} \simeq= 0.239\ GeV^{-1}$ is the $\rho\pi\gamma$
coupling
constant and $f_{\pi} \approx 93$ MeV is the pion decay constant. It is
straightforward to show that the interference of this
model-dependent amplitude with the term of order $\omega^{-1}$, namely
$\sum_{pol} {\cal M}_e {\cal M}^*_{\pi}$, vanishes identically. The reason
for this is the following: in the amplitudes that contribute to ${\cal
M}_e$, the $\rho^-$ meson couples to the spin-1 component of the
virtual $W$ gauge-boson, while the intermediate $\pi^-$ meson in the
${\cal M}_{\pi}$ amplitude is produced from the spin-0 component of the
$W$ gauge-boson. Thus, the interference of both amplitudes vanishes owing
to this orthogonality.

  Another possible model-dependent contribution arises from the  $a_1$
meson intermediate state in the process $\tau \rightarrow \nu a_1
\rightarrow \nu \gamma \rho$. The gauge-invariant amplitude
(of order $\omega$) corresponding to this process is:
\be
{\cal M}_{a_1} = \frac{-i g_{a_1 \rho\gamma} G_F g_{a_1} V_{ud}}
{ \sqrt{2} ( 2k\cdot p-m_{a_1}^2+m_\rho^2 )}
\eta^\lambda \epsilon^{*\mu} k^\sigma \epsilon_{ \mu \lambda \sigma\delta}
\left(
 -g^{\delta \phi} +\frac{q'^\delta q'^\phi}{m_{a_1}^2} 
\right) 
\bar{u} (p') O_\phi u(q) \,
\ee
where $g_{a_1}\approx 0.22$ $GeV^2$ is the $a_1$-W coupling
\cite{a1}, and $g_{a_1 \rho\gamma}\approx 0.789 e$ ($e=\sqrt{4\pi\alpha}$
in the electric charge of the proton) can be estimated from vector-meson
dominance relations using the fact that the $a_1$ meson in the orbital
excitation of the pion. We can compute the
unpolarized interference of this amplitude with the one that arises from
electric charge radiation:

\be
\sum_{pol} Re {\cal M}_{a_1} {\cal M}_e^* = \frac{4 g_{a_1\rho\gamma}e g_\rho
g_{a_1}(G_F V_{ud})^2}{( 2k\cdot p-m_{a_1}^2+m_\rho^2 )} 
k\cdot p  k\cdot q 
\left( \frac{k\cdot p}{m_\rho^2}-2  \right) \ L^2
\ee
{\it i.e.} this interference is proportional to the factor $L^2$, as in
the case of the $\rho^-$ meson decay. Therefore, the leading contribution
to the decay probability coming from model-dependent pieces vanishes in
the limit of collinear photons.

On another hand,  using the Feynman rule given in Eq. (8), we can
compute the contribution of the electric quadrupole moment to the
$\tau$ lepton decay amplitude. We obtain:

\be
{\cal M}_Q=\frac{eg_{\rho}G_F V_{ud}\gamma(0)}{2\sqrt{2} } 
\overline{u}(p') O^\alpha u(q) \left( \frac{p\cdot \epsilon}{p\cdot
k}k^{\beta} -\epsilon^{\beta} \right) \left[ k\cdot \eta g_{\alpha
\beta}+ \left( k_{\alpha}-\frac{p.k}{m_{\rho}^2} q'_{\alpha} \right)
\eta_{\beta} \right]\ ,
\ee
where $q'=p+k$.  This amplitude is gauge-invariant and of order $\omega$.
 
The interference
of the radiation amplitudes due to the electric charges, ${\cal M}_e$,  
and the electric quadrupole moment, ${\cal M}_Q$,  is given by:

\be
\sum_{pol}{\cal M}_e{\cal M}^*_Q=2( eg_\rho G_F V_{ud})^2 \gamma(0)
q\cdot k \left(-4q.k
+\frac{p.k}{m_{\rho}^2}(2[m_{\tau}^2+m_{\rho}^2-q.k]+p.k) \right) L^2\, 
\ee
{\it i.e.}, proportional to $L^2$. Thus, as in the previous case, the
interference term 2Re$\{AC^*\} \propto L^2$ will be suppressed for small
values of $\theta$.\\

  Finally, let us illustrate the effects of the model-dependent  and 
the electric quadrupole moment contributions to the distributions of
photons studied in Refs. \cite{us97,us99}. 
  In Figs. 1 and 2 we plot the angular and energy decay distributions of
photons normalized to the corresponding non-radiative decay rate 
($(1/\Gamma_{nr})d^2\Gamma/dxdy$ with $x \equiv 2\omega/M$ ($M$ is the mass of the decaying particle) and $y= \cos
\theta$) in the decays $\rho^- \rightarrow \pi^-\pi^0 \gamma$ and $\tau^-
\rightarrow \rho^- \nu\gamma$, respectively. Figure 2 includes a close 
up (right sided plot) to show the behavior of the interference terms
around $x=0.5$.

 These distributions are plotted as functions of the photon energy and for
a small value of the photon
angle emission with respect to the charged particle in the final state 
($\theta=10^0$). The short-- and long--dashed lines in both Figures 
correspond to the electric charge and the magnetic dipole
moment (taken at the canonical value $\beta(0) =2$)
contributions, respectively. Also displayed are
the curves corresponding to the model-dependent (long-short--dashed line)
and electric quadrupole moment (solid line with $\gamma(0)=2$)
contributions,  which come from their interference with the electric
charge amplitudes. We observe that, in
both cases, these additional contributions are very suppressed at small
values of photon energies and grow monotonically as $x$ 
increases until  they become  as important as the magnetic dipole moment
contribution at the end of the spectrum. In the $\rho^- \rightarrow
\pi^-\pi^0 \gamma$ decay, the
presence of the model-dependent and the electric quadrupole moment effects
reduce  the region where the 
magnetic dipole moment plays the most important role (now reduced to the
region $x \approx 0.5 \sim 0.65$). In the case of the 
$\tau^- \rightarrow V^-\nu_{\tau} \gamma$  decay, this region remains 
almost unaffected by the small contribution arising from  the electric
quadrupole moment and the $a_1$ intermediate state. However, these
contributions can be important only at the end of the photon spectrum. 

 The effects of model-dependent and electric quadrupole moment
contributions to the radiative production and decay of vector mesons may
be viewed as an obstacle for the determination of the MDM of the charged
vector mesons as suggested in
Refs. \cite{us97,us99}. However, the present analysis shows that their
contributions are suppressed for the same configurations studied in our
previous papers and that they are under good control in the kinematic
region of interest. Since model-dependent contributions are fixed 
from independent phenomenological sources, one can look at them as
background effects to be subtracted from the observable spectra. Following
the same method used to determine the triple gauge-boson vertices at the
Tevatron and LEPII colliders, measurements of the photon distributions of
our interest can provide information about one electromagnetic moment (say
the MDM) by fixing the other one to a reference (canonical) value. \\

  In conclusion, the energy spectrum of photons in the processes $V^-
\rightarrow P^- P^0\gamma$ and $\tau^- \rightarrow V^-\nu_{\tau} \gamma$
is sensitive to the effects of the magnetic dipole moment of vector
mesons $V^-$ when photons are emitted at small angles with respect to
the charged particle in the final state \cite{us97,us99}. In this paper
we have provided further support to this statement by proving  that the
model-dependent and electric quadrupole moment effects in this observable
are indeed small. We have drawn this conclusion from the fact that the
interference of the radiation amplitude off electric charges and any
other gauge-invariant amplitude is always proportional to the
Lorentz-invariant quantity $L^2$ defined in Eq. (4) (see the appendix for
a proof in the case of two-body plus photon decays; the possible
generality of this result in an arbitrary decay process is under study).
If the difficulty of reconstructing those particular kinematical
configurations
with multiple final state photons is surmount, this method would allow the
first determination of the magnetic dipole moment of charged vector
mesons.

\

{\large \bf Acknowledgements}. The authors acknowledge partial financial
support from Conacyt (Mexico) through a postdoctoral fellowship (G.T.S.)
and under contracts E-32429 and 35792-E (G.L.C.).

\

\begin{center}
{\bf Appendix}
\end{center}

  In this appendix we show that for a radiative decay $A^-(q) 
\rightarrow B^-(p) C^0(p') \gamma(\epsilon, k)$ ($q, p, p'$ and $k$ are
four-momentum vectors, and the superscripts denote the charges of
particles $A,B$ and $C$), the
interference of the
electric charge amplitude with any other gauge-invariant amplitude for
this process is proportional to the Lorentz scalar $L^2$ given in Eq.
(4). The charged particles in this reaction can be spinor, scalar o vector
particles. The generality of this property for an arbitrary
radiative process will be considered elsewhere.

  The total amplitude for the above process can be written as:
\[
{\cal M} = \epsilon^{\mu} (L_{\mu}M_0 + M_{1,\mu})\ .
\]
The first term in the r.h.s. of this equation, $L\cdot \epsilon M_0$, is
the electric charge amplitude of order $\omega^{-1}$, while $M_{1,\mu}$ is
a gauge-invariant amplitude ($k\cdot M_1=0$) starting at order $\omega$,
such that.

  The interference term between the electric charge and the remaining
amplitude, after summing over photon polarizations, is given by:
\be
{\cal I}=2Re\left [(-g^{\mu \nu})L_{\mu}\{ \sum_{pols:
A,B,C,}M_0^{*}M_{1,\nu}\}\right ]\ .
\ee
Now, the most general form of the factor within curly brackets  in this
equation is:
 \bea
V_{\mu}&=&\sum_{pols:A,B,C}M_0^{*}M_{1,\mu}\nonumber \\
&=& a_1k_{\mu}+a_2p_{\mu}+a_3q_{\mu}+\epsilon_{\mu \alpha \beta \delta} (
b_1k^{\alpha}q^{\beta}p^{\delta}+b_2p^{\alpha}k^{\beta}q^{\delta}+ 
b_3p^{\alpha}q^{\beta}k^{\delta}) \nonumber 
\eea
where $a_i,\ b_i$ are Lorentz-invariant coefficients.
   
  The gauge-invariance condition $k\cdot V=0$ imposes $a_3=-(p\cdot
k/q\cdot k)a_2$. Thus, we obtain:
\[ 
V_{\mu}=a_1k_{\mu}+a'_2L_{\mu} +\epsilon_{\mu \alpha \beta \delta} (
b_1k^{\alpha}q^{\beta}p^{\delta}+b_2p^{\alpha}k^{\beta}q^{\delta}+
b_3p^{\alpha}q^{\beta}k^{\delta})
\]

Therefore, we can complete our proof by observing that after contracting
with the four-vector $L_{\mu}$, Eq. (18) becomes:
\[
{\cal I}\sim -a'_2L^2\ .
\]

\newpage
\begin{figure}
\label{rho}
\centerline{\epsfig{file=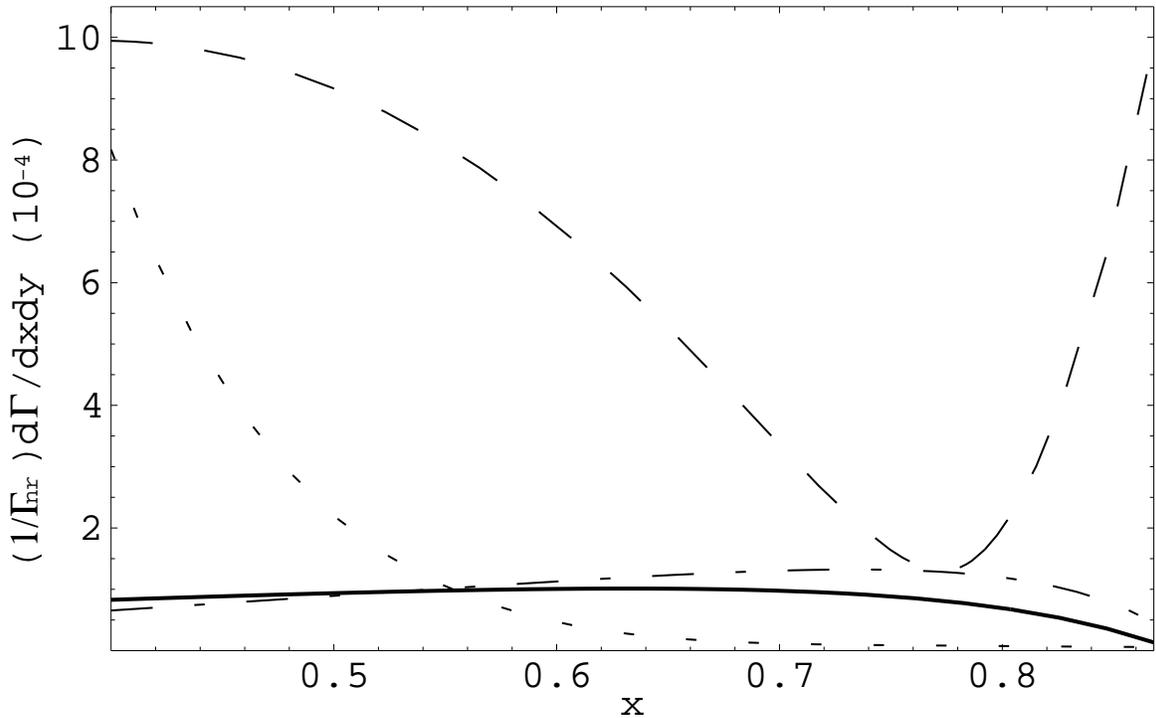,angle=0,width=6in}}
\vspace{-0.1in}
\caption{ Angular and energy decay distributions of photons in the
process $\rho^+ \rightarrow \pi^+\pi^0 \gamma$, normalized to the
non-radiative rate, as a function of the photon
energy ($x \equiv 2\omega/m_\rho$) for $\theta=10^0$ ($y
\equiv \cos\theta$). The short-- and long--dashed lines
correspond to the electric charge and the magnetic dipole moment
($\beta(0) =2$)
contributions, respectively. The long-short--dashed and solid lines
correspond to the model-dependent and the electric quadrupole moment
($\gamma(0)=2$) effects, respectively. }

\end{figure}

\newpage
\begin{figure}
\label{tauplot}
\centerline{\epsfig{file=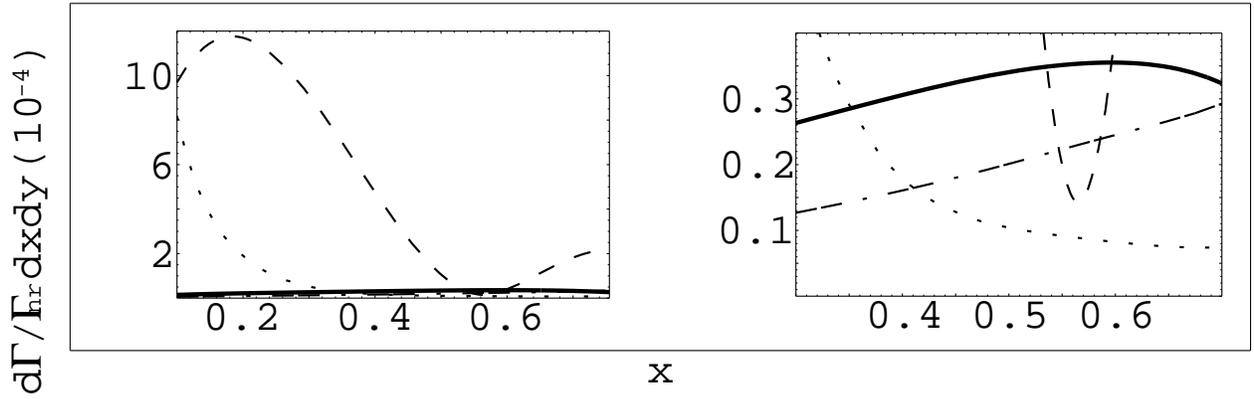,angle=0,width=6.5in}}
\vspace{-0.1in}
\caption{Same as Figure 1 for the decay $\tau^- \rightarrow \rho^-
\nu_\tau \gamma$. In this case $x \equiv 2\omega/m_\tau$ ($\omega$ is the photon energy). The solid line correspond to the
electric quadrupole moment contribution when $\gamma(0)=2$ (see text) and the
long-short-dashed line is the corresponding by $a_1$ . The figure at
the right hand side is a close up of the photon spectrum around $x=0.5$.}

\end{figure}

\end{document}